\documentclass[10pt,showpacs,aps,notitlepage,superscriptaddress,amsmath,amssymb,nofootinbib]{revtex4-1}
\usepackage{color}
\usepackage{bm}
\usepackage{times}
\usepackage{epstopdf}
\usepackage[bookmarks=true,colorlinks,linkcolor=blue,urlcolor=blue,citecolor=red]{hyperref}
\definecolor{MyDarkBlue}{rgb}{0,0.08,0.45}
\definecolor{yellow}{rgb}{0.99,0.99,0.70}
\definecolor{myback}{RGB}{204,232,207}
\definecolor{white}{rgb}{1.0,1.0,1.0}
\definecolor{black}{rgb}{0.00,0.00,0.00}

\hypersetup{
    colorlinks=true,
    linkcolor=blue,
    filecolor=magenta,
    urlcolor=cyan,
    pdftitle={Sharelatex Example},
    bookmarks=true,
    pdfpagemode=FullScreen,
}

\renewcommand\a{\alpha}
\renewcommand\b{\beta}
\renewcommand\d{\delta}

\renewcommand\l{\lambda}
\renewcommand\r{\rho}

\renewcommand\j{\psi}
\renewcommand\o{\omega}
\newcommand\e{\epsilon}
\newcommand\g{\gamma}

\newcommand\m{\mu}
\newcommand\n{\nu}
\newcommand\x{\xi}
\newcommand\p{\pi}
\newcommand\h{\theta}
\newcommand\s{\sigma}

\newcommand\w{\eta}

\renewcommand\L{\Lambda}

\renewcommand\S{\Sigma}
\renewcommand\O{\Omega}

\newcommand\D{\Delta}

\newcommand{\eq}[1]{Eq.~(\ref{#1})}

\newcommand\lb{\left(}
\newcommand\rb{\right)}
\newcommand\ls{\left[}
\newcommand\rs{\right]}
\newcommand\lc{\left\{}
\newcommand\rc{\right\}}
\newcommand{\lan}{\langle}
\newcommand{\ran}{\rangle}

\newcommand\ra{\rightarrow}

\newcommand{\non}{\nonumber\\}
\newcommand\pt{\partial}
\newcommand\mc{\mathcal}

\newcommand{\diag}{{\rm{diag}}}

\newcommand{\Tr}{{\rm Tr}}

\newcommand{\bp}{{\bm p}}
\newcommand{\bk}{{\bm k}}
\newcommand{\bq}{{\bm q}}

\newcommand{\bv}{{\bm v}}

\renewcommand{\part}{{\rm part}}

\renewcommand{\emph}[1]{{\textit{#1}}}

\begin{document}
\title{Spin Polarization Formula for Dirac Fermions at Local Equilibrium}
\author{Yu-Chen Liu}
\affiliation{Physics Department and Center for Field Theory and Particle Physics, Fudan University, Shanghai 200433, China}
\author{Xu-Guang Huang}
\email{huangxuguang@fudan.edu.cn}
\affiliation{Physics Department and Center for Field Theory and Particle Physics, Fudan University, Shanghai 200433, China}
\affiliation{Key Laboratory of Nuclear Physics and Ion-beam Application (MOE), Fudan University, Shanghai 200433, China}
\begin{abstract}
We derive a Cooper-Frye type spin polarization formula for Dirac fermions at local thermal equilibrium described by a grand canonical ensemble specified by temperature, fluid velocity, chemical potential, and spin potential. We discuss the physical meaning of different contributions to spin polarization and compare them with previous works. The present formula provides machinery to convert the spin potential computed in, e.g., relativistic spin hydrodynamics to the spin polarization observable in, e.g., heavy-ion collisions.
\end{abstract}
\maketitle

\section{Introduction}
\label{sec:intro}

The phenomena of spin polarization attract broad attention in different branches of physics from condensed matter physics to particle physics to nuclear physics. In low-energy nuclear physics, the spin polarization is an important characteristic of the nuclear equation of state~\cite{Fantoni:2001ih,Vidana:2002pc,Isayev:2003fz,Stein:2015bpa}. In heavy-ion collisions, the measurement of global spin polarization of $\L$ (and $\bar{\L}$) hyperons with respect to the direction of the reaction plane has been reported recently~\cite{STAR:2017ckg,STAR:2018gyt,STAR:2021beb} and that of $\Xi^-$ and $\O^-$ hyperons was reported in 2021~\cite{STAR:2020xbm}.

In a fluid composed of massive spin-1/2 Dirac fermions at global equilibrium, a Cooper-Frye type spin polarization formula has been derived, which shows that the mean spin vector $\bar{S}^\m$ at global equilibrium is determined solely by thermal vorticity (assuming that other external torques, like the magnetic field, do not exist)~\cite{Becattini:2013fla,Fang:2016vpj,Liu:2020flb}:
\begin{eqnarray}
\bar{S}_\mu(\bp) = -\frac{1}{8m} \e_{\m\n\a\b} p^\n \frac{\int d\Xi\cdot p\, n_F\lb 1-n_F \rb \varpi^{\a\b}}{\int d\Xi\cdot p\, n_F}\,,\;\;\;
\label{oldformula}
\end{eqnarray}
where $p^\m$ is the on-shell 4-momentum, $\Xi$ is a spacelike hypersurface on which the particles are frozen out, $n_F=n_F(p\cdot\b)$ with $n_F(x)=1/[\exp(x)+1]$ is the Fermi-Dirac function, and
\begin{eqnarray}
\varpi_{\m\n}=-\pt_{[\m}\b_{\n]}\,\;\;\;
\label{thervor}
\end{eqnarray}
is the thermal vorticity with $X_{[\m\n]}\equiv (X_{\m\n}-X_{\n\m})/2$ and $\b^\m= \b u^\m$ ($\b=1/T$ with $T$ temperature) the thermal flow velocity. The spin polarization vector is given by
\begin{eqnarray}
P_\m(\bp)=\frac{1}{s}\bar{S}_\m(\bp)=2\bar{S}_\m(\bp)\,,\;\;\;
\label{polari}
\end{eqnarray}
where $s=1/2$ is the spin quantum number. Based on this formula, the experimental data for global spin polarization of hyperons can be well reproduced from theoretical computations~\cite{Karpenko:2016jyx,Xie:2016fjj,Xie:2017upb,Li:2017slc,Shi:2017wpk,Xia:2018tes,Wei:2018zfb,Vitiuk:2019rfv,Ivanov:2019ern,Fu:2020oxj,Guo:2021uqc,Li:2021zwq,Deng:2021miw}. However, when applying the same formula to the calculation of local spin polarization (i.e., the differential spin polarization as a function of the azimuthal angle), the results deviate substantially from the data~\cite{Becattini:2017gcx, Florkowski:2019voj,Wu:2019eyi, Xie:2019jun, Xia:2019fjf,Becattini:2019ntv,Liu:2019krs,STAR:2019erd,ALICE:2021pzu}. This is a big puzzle that challenges the thermal-vorticity interpretation of spin polarization and suggests that the global equilibrium of spin degree of freedom may not be reached in real heavy-ion collisions; see more discussions in reviews~\cite{Liu:2020ymh,Becattini:2020ngo,Gao:2020vbh,Huang:2020dtn}. Possible resolutions have been proposed including admitting different spin potentials (see the definition below) to govern the amount of spin polarization~\cite{Florkowski:2019voj,Wu:2019eyi}, assuming strong chiral anomaly effects~\cite{Liu:2019krs}, and so on. New theoretical tools like spin kinetic theory~\cite{Gao:2019znl,Weickgenannt:2019dks,Hattori:2019ahi,Wang:2019moi,Liu:2020flb,Sheng:2020oqs,Guo:2020zpa,Weickgenannt:2020aaf,Yang:2020hri,Manuel:2021oah,Sheng:2021kfc,Weickgenannt:2021cuo,Lin:2021mvw} and spin hydrodynamics~\cite{Florkowski:2017ruc,Hattori:2019lfp,Montenegro:2018bcf,Florkowski:2018fap,Bhadury:2020puc,Bhadury:2020cop,Garbiso:2020puw,Gallegos:2020otk,Shi:2020htn,Fukushima:2020ucl,Li:2020eon,She:2021lhe,Gallegos:2021bzp,Hongo:2021ona,Wang:2021ngp,Peng:2021ago} that can be used to study the spin polarization out of global equilibrium are also under development though no numerical results have been reported so far.

Recently, a generalization of the spin polarization formula to local equilibrium~\footnote{We note that the term ``local equilibrium'' has been given different meanings in different context. 
In this article, we follow the convention of Refs.~\cite{Zubrev,vanWeert,Hongo:2016mqm,Becattini:2019dxo,Becattini:2018duy,Speranza:2020ilk} to use ``local equilibrium'' to name the Gibbs state described by a density operator that maximizes the von Neumann entropy under the constraint that the ensemble-averaged densities of conserved charges (energy, momentum, angular momentum, and baryon number, etc.) are fixed to the physical values. This is different from, e.g., Ref.~\cite{Hongo:2021ona}, where ``local equilibrium'' is used to name the local state in which the strict hydrodynamic description can apply.} has been proposed in which not only the thermal vorticity but also the thermal shear tensor contributes to the mean spin vector~\cite{Becattini:2021suc,Liu:2021uhn}. It was found that the thermal-shear contribution could be an important component for local spin polarization under certain conditions: For example, for hyperons that are isothermally frozen out~\cite{Becattini:2021iol}, or for hyperons whose spin polarization is dominantly determined by its strange quark constituents~\cite{Fu:2021pok}, the calculated spin polarization can match the experimental data. But it was also argued that the results could be sensitive to the equation of state of the matter~\cite{Yi:2021ryh} and to the procedure of how the thermal vorticity and thermal shear contributions are combined~\cite{Florkowski:2021xvy}.

It is known that the description of local equilibrium state is not unique; it depends on the choices of the energy-momentum tensor and spin current in a compatible way~\cite{Becattini:2018duy,Speranza:2020ilk,Tinti:2020gyh}. For example, when choosing the Belinfante symmetric energy-momentum tensor for Dirac fermions, the corresponding spin current vanishes, and the local equilibrium is specified by local thermal flow vector $\b^\m$. 
On the other hand, if the canonical form for the energy-momentum tensor and spin current are chosen, the local equilibrium state admits a new, independent, rank-two tensor $\o^{\m\n}$ (called spin potential, which is conjugate to local canonical spin density) in addition to $\b^\m$~\footnote{Note that, generically, the $\b^\m$ in this case does not coincide with the $\b^\m$ in the Belinfante case.}. The previous studies did not consider the contribution of $\o^{\m\n}$ to the spin polarization formula though the canonical spin current was adopted to describe the local equilibrium state~\cite{Becattini:2021suc,Liu:2021uhn,Becattini:2021iol,Fu:2021pok,Yi:2021ryh}. 
In this article, we will re-visit the spin polarization at local equilibrium and derive a new spin polarization formula by using the canonical forms of energy-momentum and spin current tensors. We will show that the spin potential makes an additional contribution in the new spin polarization formula and we will discuss the physical content of the new formula.

We will take metric $\w_{\m\n}=\diag(1, -1, -1, -1)$ and the natural units $k_B=c=1$. The Levi-Civita tensor $\e^{\m\n\r\s}$ is normalized as $\e^{0123}=-\e_{0123}=1$.

\section{Derivation using kinetic theory}
\label{sec:ckt}
We recall that for spin-1/2 Dirac fermions the canonical spin current operator is given by (we restore $\hbar$ in this section)
\begin{eqnarray}
\begin{split}
\hat{S}^{\m,\r\s} (x) &\equiv \frac{\hbar}{4}{\bar{\hat\j}}(x) \lc \g^\m, \s^{\r\s} \rc \hat{\j}(x)\\
&= \int \frac{d^4 p}{(2\p)^4} \hat{S}^{\m,\r\s} (x,p)\,,
\label{canospin}
\end{split}
\end{eqnarray}
where $\s^{\m\n}=i[\g^\m, \g^\n]/2$ and $\hat{S}^{\m,\r\s} (x,p)$ is the canonical spin current operator in phase space which is related to Wigner operator as follows,
\begin{eqnarray}
\label{spinwigner}
\hat{S}^{\m,\r\s} (x,p) &=&\frac{\hbar}{4} \Tr_{\rm D} \ls \lc \g^\m, \s^{\r\s} \rc \hat{W}(x,p) \rs\,,\\
\label{Wigneroperator}
\hat{W}(x,p)&=&\int d^4s  e^{-ip\cdot s/\hbar} \bar{\hat\j}\left(x+\frac{s}{2}\right) \otimes \hat{\j} \left(x-\frac{s}{2}\right),\;\;\;\;
\end{eqnarray}
where $\Tr_{\rm D}$ is trace over Dirac space and $[\bar{\hat\psi}\otimes\hat{\psi}]_{ab}\equiv\bar{\hat\psi}_b\hat{\psi}_{a}$ with $a,b$ spinor indices. The Wigner function is defined as the ensemble average of the Wigner operator: $W(x,p)= \lan \hat{W}(x,p) \ran = {\rm Tr} \ls \hat{\r}\hat{W}(x,p) \rs$ with $\hat{\r}$ the density operator and ${\rm Tr}$ the trace over a complete set of microstates of the system. For free Dirac fermions at local equilibrium, $\hat{\r}=\hat{\r}_{\rm LE}$, the Wigner function can be directly calculated order by order in $\hbar$ using second quantization, as we will see in next section. In this section, however, we will use the results from chiral kinetic theory and spin kinetic theory which provide an alternative way to calculate the Wigner function. First, we decompose the Wigner function on the basis of Clifford algebra:
\begin{equation}
\label{clifford}
W=\frac{1}{4}[\mathcal{F}+i\g^5\mathcal{P}+\g^\m\mathcal{V}_\m
+\g^5\g^\m\mathcal{A}_\m+\frac{1}{2}\s^{\m\n}\mathcal{S}_{\m\n}]
\end{equation}
with all the Clifford coefficients being real. Further, one can find that
\begin{equation}
S^{\mu,\rho\s}(x,p)  = -\frac{\hbar}{2} \e^{\mu\rho\s\lambda}\mc{A}_\l(x,p) \,,\\
\end{equation}
where $S^{\m,\r\s}(x,p)=\lan\hat{S}^{\m,\r\s}(x,p)\ran$. The above relation identifies the canonical spin vector $S^\m$ with the axial current $\mc{A}^\m$ in phase space:
\begin{equation}
\label{relationsa}
S^\m(x,p)=-\frac{1}{6}\e^{\m\n\r\s}S_{\n,\r\s}(x,p)=\frac{\hbar}{2}\mc{A}^\m(x,p).
\end{equation}
The dynamic equation of Wigner function reads (see, e.g., Refs.~\cite{Liu:2020flb,Gao:2019znl,Weickgenannt:2019dks,Hattori:2019ahi,Wang:2019moi,Sheng:2020oqs})
\begin{eqnarray}
\ls \g^\m\lb\frac{i\hbar}{2}\pt_\m + p_\m \rb - m \rs W = 0\,,
\label{dynamicofwigner}
\end{eqnarray}
from which one can solve for $\mc{A}_\m$ and other Clifford coefficients order by order in $\hbar$.

\subsection{Massless case}
\label{sec:massless}
Let us first consider the massless limit, $m=0$. The solution for $\mc{A}_\m$  up to $O(\hbar)$ is (see, e.g., Refs.~\cite{Hidaka:2016yjf,Huang:2018wdl,Liu:2018xip,Gao:2018wmr} for more details)
\begin{equation}
\begin{split}
\mc{A}_\m &= 4\p \d(p^2) \lb p_\m f_5 + \hbar \S^n_{\m\n} \pt^\n f \rb,
\label{eq:Amassless}
\end{split}
\end{equation}
where $f_5=f_R-f_L$ and $f=f_R+f_L$ are the axial and vector distribution functions with $f_R$ and $f_L$ the distribution functions of right-handed and left-handed fermions, respectively, $\S_n^{\m\n}\equiv \e^{\m\n\r\s}p_\r n_\s/(2p\cdot n)$ represents the single-particle spin tensor with $n^\m$ a normalized timelike frame-fixing vector, $n^2=1$. 
Similarly, the vector current reads
\begin{equation}
\begin{split}
 \mc{V}_\m= 4\p \d(p^2) \lb p_\m f + \hbar\S^n_{\m\n}\pt^\n f_5 \rb.
 \label{eq:Vmassless}
\end{split}
\end{equation}
The concrete forms of the distribution functions $f$ and $f_5$ cannot be determined without knowing the collision terms. However, for collisional processes that conserve energy, momentum, particle numbers, and angular momentum, the local-equilibrium distributions should depend only on the linear combination of these quantities. Therefore, we assume the distribution functions of right-handed and left-handed fermions to be of Fermi-Dirac form and read
\begin{eqnarray}
\label{masslessdis}
f_{R/L} &=&\h(p_0)n_F(g_{R/L})+\h(-p_0)[1-n_F(g_{R/L})]\,,
\label{LEchiraldistribution}
\end{eqnarray}
where $g_{R/L}=\a+p\cdot \b \mp \frac{\hbar}{2} \S_n^{\m\n}\o_{\m\n}$, $n_F(x)=(e^x+1)^{-1}$ is the Fermi-Dirac function, and the Lagrangian multipliers  $\b_\m, \a$, $\o_{\m\n}$ have the following physical meanings: $\o_{\m\n}=-\o_{\n\m}$ is the spin potential tensor~\footnote{Note that $\o^{\m\n}$ is transverse to $n^\m$ so that it contains only three independent components in accordance with the three components of spin vector.}, $\b_\m$ is the thermal flow velocity, and $\a=-\b\m$ with $\m$ the chemical potential. We will regard $\b^\m, \a$, and $\o_{\m\n}$ as $O(1)$ in $\hbar$ power counting. Substituting $f_{R/L}$ into $\mc{A}_\m$ and keeping terms up to $O(\hbar)$, we obtain
\begin{equation}
\begin{split}
 \mc{A}_\m
&=-8\p \d(p^2) \bigg\{\frac{\hbar}{4} \e_{\m\n\a\b} p^\n \o^{\a\b}+ \hbar \S^n_{\m\n} \ls p_\l (\x^{\n\l}+\D\o^{\n\l} )+ \pt^\n \a \rs \bigg\}\\&\quad\times [\h(p_0)-\h(-p_0)] n_F\lb 1-n_F \rb ,
\label{spinpolmassless}
\end{split}
\end{equation}
where $\xi^{\n\l}=\pt^{(\n}\b^{\l)}$ is the thermal shear tensor, $\D\o^{\n\l}=\o^{\n\l}-\varpi^{\n\l}$ is the deviation of spin potential from thermal vorticity, and we have used the abbreviation $n_F=n_F(\a+p\cdot\b)$. The first term is the spin polarization induced by spin potential, which is not necessarily equal to thermal vorticity at local equilibrium. 
The second term is the spin polarization induced by the thermal shear tensor and the deviation of spin potential from the thermal vorticity. The last term is due to the spatial inhomogeneity of $\a$.

\subsection{Massive case}
\label{sec:massive}
Next, we consider massive fermions. Unlike the massless fermion whose spin is slaved to the momentum, the spin of a massive fermion is an independent degree of freedom. This makes the kinetic description of the spin dynamics of massive fermion more complicated: Comparing to the massless case where two kinetic equations are enough, we need four kinetic equations (one for the particle number and three for the spin vector) in the massive case~\cite{Gao:2019znl,Weickgenannt:2019dks,Hattori:2019ahi,Wang:2019moi,Liu:2020flb,Sheng:2020oqs}. Instead of presenting all these kinetic equations, we here list a few relations for the Clifford coefficients in Eq.~\eqref{clifford} which can be obtained through the $\hbar$ expansion of Eq.~\eqref{dynamicofwigner} and will be enough for our purpose:
\begin{eqnarray}
\label{skteqs}
\begin{split}
\mc{F}&=4\p m\d(p^2-m^2)f +O(\hbar^2), \\
\mc{S}^{\m\n} &= 4\p m\d(p^2-m^2) {\mc{M}}^{\m\n} + O(\hbar^2), \\
\hbar\pt_\m \mc{F}&= 2 p^\n \mc{S}_{\m\n} +  O(\hbar^2), \\
\mc{A}_\m &=- \frac{1}{2m} \e_{\m\n\r\s}p^\n{\mc{S}}^{\r\s} + O(\hbar^2),\\
\mc{V}_\m &= 4\p \d(p^2-m^2) p_\m f +\frac{\hbar}{2m}\pt^\n\mc{S}_{\m\n} +O(\hbar^2),
\end{split}
\end{eqnarray}
where $\mc{F}$ and $\mc{S}^{\m\n}$ are parameterized by $f$ and $\mc{M}^{\m\n}$ which are identified as the on-shell particle number and tensorial spin densities (distribution functions) in phase space. We count $f$ as $O(1)$ and $\mc{M}^{\m\n}$ as $O(\hbar)$ quantities.
Multiplying the second equation in \eqref{skteqs} by $p\cdot n$, then substituting $\mc{S}^{\m\n}$ into the forth equation, and using the first and third equations and the Schouten identity $p_\l \e_{\m\n\r\s} + p_\m \e_{\n\r\s\l} +p_\n \e_{\r\s\l\m}+p_\r \e_{\s\l\m\n}+p_\s \e_{\l\m\n\r}=0$, we express $\mc{A}^\m$ in terms of $f$ and $\mc{M}^{\m\n}$, up to $O(\hbar)$~\footnote{This equation has been used in Ref.~\cite{Sheng:2020oqs} to show that in the massless limit, $m\ra 0$, $\mc{A}^\m$ smoothly goes into its massless counterpart in Eq.~\eqref{eq:Amassless} by recognizing $\S^n_{\r\s} {\mc{M}}^{\r\s}$ as the axial distribution function $f_5$.},
\begin{equation}
\begin{split}
\mc{A}_\m&=4\p \d(p^2-m^2) \bigg( p_\m \S^n_{\r\s} {\mc{M}}^{\r\s} + \hbar \S^n_{\m\n} \pt^\n f  - \frac{m^2}{2 p\cdot n} \e_{\m\n\r\s} n^\n {\mc{M}}^{\r\s} \bigg).
\label{Amassivetomless}
\end{split}
\end{equation}
This expression shows that only the $n^\m$-transverse components of $\mc{M}^{\m\n}$ (corresponding to the three components of spin vector) contribute to $\mc{A}^\mu$ at $O(\hbar)$ order. In fact, by decomposing $\mc{M}^{\m\n}=\mc{E}^\m n^\n-\mc{E}^\n n^\m+\e^{\m\n\a\b}n_\a \mc{M}_\b$, it can be shown that $\mc{E}^\m$ is determined by $f$ and $\mc{M}^\m$~\cite{Sheng:2020oqs}. Using once again the Schouten identity for the first term in Eq.~\eqref{Amassivetomless}, one can re-write $\mc{A}^\m$ as
\begin{equation}
\begin{split}
 \mc{A}_\m
&=-8\p \d(p^2-m^2) \bigg( \frac{1}{4} \e_{\m\n\a\b} p^\n \mc{M}^{\a\b}_\perp-\frac{\hbar}{2} \S^n_{\m\n} \pt^\n f +  \S^n_{\m\n} \mc{M}_\perp^{\n\l}p_\l \bigg ),
\label{spinpolmassless2}
\end{split}
\end{equation}
where $\mc{M}^{\m\n}_\perp=\e^{\m\n\a\b}n_\a \mc{M}_\b$.
At local equilibrium, for an arbitrary space-like direction $r^\m$ (with $r^2=-1$) regarded as the spin quantization axis, both $f$ and $\mc{M}_\perp^{\m\n} r_{\m\n}$ (the tensorial spin density projected onto $r^\m$) with $r_{\m\n}=\e_{\m\n\r\s}n^\r r^\s/2$ should be determined by $\a, \beta$, and $\o^{\m\n}$ via $f=f_++f_-$ and $\mc{M}_\perp^{\m\n} r_{\m\n}=f_+-f_-$ where
\begin{eqnarray}
\label{ltemassive}
f_{\pm} &=&\h(p_0)n_F(g_{\pm})+\h(-p_0)[1-n_F(g_{\pm})]\;,
\end{eqnarray}
with $g_\pm=\a+p\cdot\b\mp\frac{\hbar}{2}\o^{\a\b}r_{\a\b}$. We have used the same assumption as that for Eq.~\eqref{LEchiraldistribution} to obtain the above equation. One thus finds that $f=2[ \h(p_0)n_F+\h(-p_0)(1-n_F)]+O(\hbar^2)$ with $n_F=n_F (\a+p\cdot\b)$ and $\mc{M}_\perp^{\m\n}=-\hbar\o^{\m\n} [\h(p_0)-\h(-p_0)]n_F'(\a+p\cdot\b)+O(\hbar^2)$. Substituting $f$ and $\mc{M}_\perp^{\m\n}$ into Eq.~\eqref{spinpolmassless2}, we obtain
\begin{equation}
\begin{split}
 \mc{A}_\m
&=-8\p \hbar \d(p^2-m^2)  \bigg\{ \frac{1}{4} \e_{\m\n\a\b} p^\n \o^{\a\b} + \S^n_{\m\n} \ls p_\l ( \x^{\n\l}+\D\o^{\n\l} )+ \pt^\n \a \rs \bigg\}\\&\quad\times[\h(p_0)-\h(-p_0)] n_F\lb 1-n_F \rb.
\label{spinpolmassive}
\end{split}
\end{equation}
Note that Eq.~\eqref{spinpolmassive} is almost identical with its massless counterpart, Eq.~\eqref{spinpolmassless}, except for the different on-shell condition.

\section{Derivation using local equilibrium density operator}
\label{sec:NEDO}
In this section, we present another derivation of Eq.~\eqref{spinpolmassive} by using the local equilibrium density operator~\cite{Zubrev,vanWeert}. We will focus on massive Dirac fermions and we will follow closely Refs.~\cite{Becattini:2021suc,Becattini:2020sww}. We set $\hbar=1$ in this section. The starting point is the following density operator describing a local equilibrium state specified by $\a(x), \b_\m(x)$, and $\o_{\m\n}(x)$,
\begin{equation}
\begin{split}
\label{rhoLE}
\hat{\r}_{\rm LE} &= \frac{1}{Z_{\rm LE}} \exp \bigg\{ -\int d\Xi_\m(y) \big[ \hat{T}^{\m\n}(y)\b_\n(y) +\hat{J}^\m(y)\a(y) - \frac{1}{2}\hat{S}^{\m,\r\s}(y)\o_{\r\s}(y) \big] \bigg\},
\end{split}
\end{equation}
where $\hat{T}^{\m\n}$, $\hat{J}^\m$, and $\hat{S}^{\m,\r\s}$ are the canonical operators for energy-momentum tensor, vector current, and spin current, respectively, and $Z_{\rm LE}$ is the partition function. For a local Heisenberg operator $\hat{O}(x)$, we have its ensemble average given by
\begin{eqnarray}
\label{ensembleave}
O(x)&\equiv&\lan\hat{O}(x)\ran\equiv \Tr[\hat{\r}_{\rm LE}\hat{O}(x)]
 = \frac{1}{Z_{\rm LE}}\Tr\ls e^{\hat{A}+\hat{B}}\hat{O}(x)\rs,\qquad
\end{eqnarray}
where we introduced the abbreviations
\begin{eqnarray}
\hat{A}&=&-\hat{P}^\m\b_\m(x)-\hat{N}\a(x),\\
\hat{B}&=&-\int d\Xi_\m(y) \big[ \hat{T}^{\m\n}(y)\D\b_\n(y)+\hat{J}^\m(y)\D\a(y)-\frac{1}{2}\hat{S}^{\m,\r\s}(y)\o_{\r\s}(y)\big],
\end{eqnarray}
with $\hat{P}^\m=\int d\Xi_\n(y)\hat{T}^{\n\m}(y)$, $\hat{N}=\int d\Xi_\n(y)\hat{J}^{\n}(y)$, $\D\b_\m(y)=\b_\m(y)-\b_\m(x)$, and $\D\a(y)=\a(y)-\a(x)$.

In the following, we assume that $\o_{\r\s}(y)$ is small on hypersurface $\Xi$ and $\D\b_\m(y)$ and $\D\a(y)$ are also small on $\Xi$. This is justified when $\b_\m$ and $\a$ vary slowly within the correlation lengths between $\hat{O}$ and the operators $\hat{T}^{\m\n}$ and $\hat{J}^\m$. Thus Eq.~\eqref{ensembleave} can be calculated in a perturbative way in derivative expansion using the power counting rule $\D\b_\n(y)=(y-x)^\l\pt_\l\b_\n(x)+\cdots=O(\pt)$, $\D\a(y)=(y-x)^\l\pt_\l\a(x)+\cdots=O(\pt)$, and $\o_{\r\s}(y)=O(\pt)$. It should be emphasized that in the current content the gradient expansion is equivalent to the $\hbar$ expansion that we adopted in last section. Adopting $
e^{\hat{A}+\hat{B}}= e^{\hat{A}}+ e^{\hat{A}} \int^1_0 e^{-\l \hat{A}} \hat{B}e^{\l \hat{A}} + \cdot\cdot\cdot $,  we expand the ensemble average $O(x)$ into the following form
\begin{eqnarray}
\label{ensembleave3}
O(x) &=& O_0(x) + O_1(x)+\cdots\,,
\end{eqnarray}
where
\begin{eqnarray}
\label{ensembleave4}
O_0(x) = \langle \hat{O}(x) \rangle_{0} \equiv \frac{1}{Z_0}\Tr\lb e^{\hat{A}}\hat{O}(x)\rb,\qquad\\
O_1(x) \equiv \lan\hat{O}(x)\ran_{(T)} + \lan\hat{O}(x)\ran_{(J)} + \lan\hat{O}(x)\ran_{(S)}\,\,,
\end{eqnarray}
with
\begin{equation}
\begin{split}
\lan\hat{O}(x)\ran_{(T)} &\equiv - \int_0^1d\l \int d\Xi_{\m}(y)\D\b_{\n}(y)\lan\hat{T}^{\m\n}(y-i\l \b(x))\hat{O}(x)\ran_{0,c}\,,\quad\\
\lan\hat{O}(x)\ran_{(S)} &\equiv  \frac{1}{2}\int_0^1d\l \int d\Xi_{\m}(y)\o_{\r\s}(y)\lan\hat{S}^{\m,\r\s}(y-i\l \b(x))\hat{O}(x)\ran_{0,c}\,,\qquad\\
\lan\hat{O}(x)\ran_{(J)} &\equiv - \int_0^1d\l \int d\Xi_{\m}(y)\D\a(y)\lan\hat{J}^{\m}(y-i\l \b(x))\hat{O}(x)\ran_{0,c}\,,
\label{OTJS}
\end{split}
\end{equation}
and $Z_0=\Tr e^{\hat{A}}$. In the above expressions, $\lan\cdots\ran_{0,c}$ means the connected part of the correlator.

We now apply the above procedure to the Wigner function for massive Dirac fermions:
\begin{equation}
W(x,p)=W_0(x,p)+ W_1(x,p)+ \cdots.
\end{equation}
We first express the Wigner operator in Eq.~\eqref{Wigneroperator} using the free field operator
\begin{equation}
\begin{split}
\hat{\j}(x)&=\sum^2_{\s=1}\frac{1}{(2\p)^{3/2}} \int \frac{d^3\bk}{2E_k} \big[ u_\s(\bk) e^{-ik\cdot x} \hat{a}_\s(\bk) +  v_\s(\bk) e^{ik\cdot x} \hat{b}^\dag_\s(\bk) \big]\,,
\end{split}
\end{equation}
where $E_k=\sqrt{\bk^2+m^2}$, the creation and annihilation operators for particles and antiparticles satisfy the anti-commutation relation $\{ \hat{a}_\s(\bk), \hat{a}^\dag_{\s'}(\bq) \}=\{ \hat{b}_\s(\bk), \hat{b}^\dag_{\s'}(\bq) \} = 2E_k \d_{\s\s'}\d^3(\bk-\bm{q})$, and the spinors satisfy the relation $\bar{u}_\s(\bk) u_{\s'}(\bk)=2m\d_{\s\s'}=-\bar{v}_\s(\bk) v_{\s'}(\bk)$. Substituting the above equation into Eq.~\eqref{Wigneroperator}, we arrive at the following form for the Wigner operator (Here and below, we focus on particle branch only):
\begin{equation}
\begin{split}
\hat{W}(x,p)&= 2\p\sum^2_{\s,\s'=1} \int \frac{d^3\bq}{2E_q} \int \frac{d^3\bk}{2E_k} e^{i(k-q)\cdot x}\d^4\lb p-\frac{k}{2}-\frac{q}{2}\rb\bar{u}_\s(\bk)\otimes u_{\s'}(\bq) \hat{a}^\dag_\s(\bk) \hat{a}_{\s'}(\bq).
\label{wigneropfreefield}
\end{split}
\end{equation}

With the above preparations, we now calculate the Wigner function. Using $\lan \hat{a}^\dag_\s(\bk) \hat{a}_{\s'}(\bq)\ran_0=2E_k\d_{\s\s'}\d^3(\bk-\bq) n_F(\a+p\cdot\b)$, one find that the zeroth order Wigner function takes the form $W_0(x,p)=2\p (p\!\!\!/+m)\h(p_0)\d(p^2-m^2)n_F(\a+p\cdot\b)$, which is spin independent: $\Tr_D \ls \g^\m\g_5 W_0(x,p)\rs = 0$. This gives the leading-order vector current as
\begin{eqnarray}
\label{Wigvec}
\mc{V_\m}=8\p \d(p^2-m^2) \h(p_0)p_\m n_F(\a+p\cdot\b),
\end{eqnarray}
which coincides with the last expression in \eq{skteqs} at local equilibrium with particle branch picked up. The first order Wigner function can be derived by using Eq.~\eqref{OTJS}, with the help of
$\hat{T}^{\m\n}(x)=\int \frac{d^4 p}{(2\p)^4} p^\n \Tr_D \ls \g^\m \hat{W}(x,p) \rs$,
$\hat{J}^\m (x) = \int \frac{d^4 p}{(2\p)^4} \Tr_D \ls \g^\m \hat{W}(x,p) \rs$, and Eq.~\eqref{spinwigner}. Substituting the Wigner operator in Eq.~\eqref{wigneropfreefield} into the right-hand side of Eq.~\eqref{OTJS}, we obtain
\begin{eqnarray}
\label{Wtsj}
\lan \hat{W}(x,p)\ran_{(T/S/J)}
&=& (2\p)\int_0^1 d\l \int \frac{d^3\bk}{2E_k} \int \frac{d^3\bq}{2E_q}
 \d^4\lb p-\frac{q+k}{2} \rb (\g\cdot k+m)\hat{t}_\m I_{(T/S/J)}^\m (\g\cdot q+m) \non
&&\times e^{\l(k-q)\cdot\b(x)} n_F(k)\ls 1 - n_F(q) \rs\,,
\end{eqnarray}
where $\hat{t}^\m$ is the normal vector of the hypersurface $\Xi$, $n_F(p)=n_F(\a+\b(x)\cdot p)$, 
and
\begin{equation}
\begin{split}
I_{(T)}^\m &\equiv - \g^\m p^\n\ls\pt_\l\b_\n(x)\rs\D^\l_\b \ls i\pt_q^\b \d^3(\bm{q}-\bm{k}) \rs\,,\\
I_{(S)}^\m & \equiv \frac{1}{4}\e^{\m\n\r\s} \g^5\g_\n \o_{\r\s} \d^3(\bm{q}-\bm{k})\,, \\
I_{(J)}^\m &\equiv \g^\m \ls\pt_\l \a(x)\rs \D^\l_\b \ls i\pt_q^\b \d^3(\bm{q}-\bm{k}) \rs\,,
\end{split}
\end{equation}
with $\D^{\m\n}=\w^{\m\n}-\hat{t}^\m\hat{t}^\n$. 
In order to derive Eq.~\eqref{Wtsj}, we have assumed that $\hat{t}_\m$ does not deviate strongly from the coordinate time direction so that approximately
\begin{equation}
\begin{split}
\int d\Xi_{\m}(y)(y-x)^{\a}e^{-i(p-q)\cdot(y-x)}\approx (2\p)^3
\hat{t}_\m \D^{\a}_{\b}\frac{i\pt}{\pt p_{\b}}\d^{3}(\bp-\bq)+(2\p)^3\hat{t}\cdot(y-x)\hat{t}^\a\hat{t}_\m\d^{3}(\bp-\bq).
\end{split}
\end{equation}
In heavy-ion collisions, the true freeze-out hypersurface $\Xi$ may not satisfy the above condition and in this case one can choose $\hat{t}^\m$ as the time direction of an earlier-stage fluid occupying a flat three-dimensional hypersurface $\Xi_B$, if the integration over the four-volume encompassed by $\Xi$ and $\Xi_B$ does not contribute to the spin vector~\cite{Becattini:2021suc}.

With the first order solution of Wigner function in Eq.~\eqref{Wtsj}, we directly derive the following axial vector following similar procedures as Ref.~\cite{Becattini:2021suc}:
\begin{equation}
\begin{split}
\label{doa}
\mc{A}_\m &=- 8\p \d( p^2-m^2 )\h(p_0) n_F ( 1 - n_F ) \bigg\{ \frac{1}{4} \e_{\m\n\r\s} p^\n \o^{\r\s}+  \S^{\hat{t}}_{\m\n} \ls (\x^{\n\l}+\D\o^{\n\l}) p_\l + \pt^\n \a \rs \bigg\},
\end{split}
\end{equation}
which is exactly the particle branch of Eq.~\eqref{spinpolmassive} with $n^\m = \hat{t}^\m$. The anti-particle branch can be similarly obtained by replacing \eq{wigneropfreefield} with the anti-particle Wigner operator.

\section{A spin Cooper-Frye formula}
\label{sec:meanspin}
With having $\mc{A}^\m(x,p)$, the spin vector $S^\m(x,p)$ is given by \eq{relationsa}. Let us focus on the spin polarization of particles; the anti-particle spin polarization can be similarly discussed. Integrating \eq{spinpolmassless} or Eq.~\eqref{spinpolmassive} [or \eq{doa}] over $p_0$ from $0$ to $\infty$ we obtain the on-shell spin vector (we set $\hbar=1$):
\begin{eqnarray}
\label{sphasespace}
S^\m (x,\bp)&=&\frac{1}{2}\int_0^\infty\frac{dp_0}{2\p}\mc{A}^\m(x,p)\non
&=&- \frac{1}{4E_p}\lc \e^{\m\n\r\s} p_\n \o_{\r\s}+  4\S_{n}^{\m\n} \ls (\x_{\n\l}+\D\o_{\n\l}) p^\l + \pt_\n \a \rs \rc n_F( 1 - n_F ) ,
\end{eqnarray}
where, on the right-hand side, $p_0=E_p=\sqrt{\bp^2+m^2}$ for both $m>0$ and $m=0$ cases. In order to more clearly understand the different terms in \eq{sphasespace}, it is instructive to choose $n^\m$ as the coordinate time direction, i.e., $n^\m=(1,0,0,0)$, and consider the low-velocity limit, i.e., $u^\m\approx(1,\bv)$ with $|\bv|\ll 1$. The zeroth component of the spin vector is (half) the helicity induced by finite spin potential:
\begin{eqnarray}
\label{zerothcomp}
S^0(x,\bp)\approx\frac{1}{2}\ls n_F\lb\a+\b E_p-\frac{\bm\o\cdot\bp}{2E_p}\rb-n_F\lb\a+\b E_p+\frac{\bm\o\cdot\bp}{2E_p}\rb\rs,
\end{eqnarray}
where $\o^i=\e^{ijk}\o_{jk}/2$ is the spin potential vector. The spatial components are $\bm S=\bm S_{(\o)}+\bm S_{(\O)}+\bm S_{(\s)}+\bm S_{(T)}+\bm S_{(\a)}$ with
\begin{eqnarray}
\label{firstspa}
S^i_{(\o)}(x,\bp)&=&\ls \frac{\o^i}{2}-\frac{\bp^2\o^i-\bm\o\cdot\bp\, p^i}{2E_p^2}\rs n_F(1-n_F),\\
S^i_{(\O)}(x,\bp)&=&\frac{\bp^2\O^i-\bm\O\cdot\bp\, p^i}{2T\,E_p^2} n_F(1-n_F),\\
S^i_{(\s)}(x,\bp)&=&\frac{\e^{ijk}\,p^j p^l\,\s_{kl}}{2T\,E_p^2} n_F(1-n_F),\\
S^i_{(T)}(x,\bp)&=&-\frac{\lb\bp\times\bm\nabla T\rb^i}{2T^2\,E_p}\, n_F(1-n_F),\\
S^i_{(\a)}(x,\bp)&=&\frac{\lb\bp\times\bm\nabla\a\rb^i}{2E_p^2}n_F(1-n_F),
\end{eqnarray}
where $\bm\O=(\bm\nabla\times\bv)/2$ is the non-relativistic vorticity, $\s_{ij}=\lb\pt_i v_j+\pt_j v_i+2\d_{ij}\bm\nabla\cdot\bv/3\rb/2$ is the non-relativistic shear tensor, $n_F=n_F(\a+\b E_p)$, and we have kept only the leading terms in $\bv$.

Equation (\ref{sphasespace}) is particularly useful for the calculation of spin polarization in transport models because the transport models can provide full phase-space information about the particles' production and evolution. But it is less useful for hydrodynamic modeling of heavy-ion collisions in which the particles are produced (emitted) on some freeze-out hypersurface $\Xi$. In this case, we need to average $S^\m(x,\bp)$ over $\Xi$ to obtain the mean spin vector~\footnote{This definition of the mean spin vector can also be expressed as $\bar{S}^\m=-\e^{\m\n\r\s}p_\n\bar{S}_{\r\s}/(2E_p)$ with
$\bar{S}_{\r\s}$ the mean spin tensor given by $\bar{S}^{\m\n}=\int_0^\infty dp_0\int d\Xi_\l S^{\l,\m\n}(x,p)/\ls\int_0^\infty dp_0 \int d\Xi_\l  \mc{V}^\l(x,p)\rs$. We note that this is different from Refs.~\cite{Becattini:2013fla,Becattini:2021suc,Becattini:2020sww} where the mean spin vector is defined as $\bar{S}^\m=-\e^{\m\n\r\s}p_\n\bar{S}_{\r\s}/(2m)$ which works only for massive fermions. With this latter definition, finally $\bar{S}^\m$ is also expressed by \eq{newmeanspin} but with $E_p$ appearing in the denominator replaced by $m$.}
\begin{eqnarray}
\label{newmeanspin}
\bar{S}^{\m}(\bp)&=&\frac{\int d\Xi\cdot p \; S^{\m}(x,\bp)}{2\int d\Xi\cdot p\; n_F}\non
&=& - \frac{\int d\Xi\cdot p \lc \e_{\m\n\a\b} p^\n \o^{\a\b}
+ 4\S^n_{\m\n} \ls p_\l (\x^{\n\l}+\D\o^{\n\l})+ \pt^\n \a \rs \rc n_F ( 1 - n_F )}{8E_p\;\int d\Xi\cdot p\; n_F},
\end{eqnarray}
where $p^\m$ on the right-hand side is on-shell. This is a Cooper-Frye type formula for spin polarization which connects the momentum-space distribution of mean spin vector of the particles emitted from $\Xi$ with the fluid properties characterized by $\o^{\m\n}(x)$, $\b^\m(x)$, and $\a(x)$ on $\Xi$. In the following, we give some comments on formula (\ref{newmeanspin}).

(1) At global equilibrium, $\x^{\m\n}=\pt^\m \a = 0$ and $\o^{\m\n}=\varpi^{\m\n}$~\cite{Liu:2018xip,Liu:2020flb,Becattini:2012tc}, Eq.~\eqref{newmeanspin} becomes
\begin{eqnarray}
\bar{S}^\m (\bp) &=& - \frac{\e_{\m\n\a\b}\,p^\n \int d\Xi\cdot p\; \varpi^{\a\b}\;n_F ( 1 - n_F )}{8E_p\int d\Xi\cdot p\; n_F}  ,
\label{oldmeanspin}
\end{eqnarray}
which is essentially Eq.~\eqref{oldformula} when applied to massive fermions.

(2) There is a thermal shear contribution to $\bar{S}^\m (\bp)$ which may have important implications to phenomenology of spin polarization in heavy-ion collisions, as having been discussed in Ref.~\cite{Becattini:2021suc} with $n^\m=\hat{t}^\m$ and in Ref.~\cite{Liu:2021uhn} with $n^\m=u^\m$.

(3) When the system permits finite helicity density at local equilibrium, there will appear a new contribution to the mean spin vector. This can be derived, for massless fermions, by relaxing $g_{R/L}$ in \eq{masslessdis} to $g_{R/L}=\a_{R/L}+p\cdot \b \mp \frac{\hbar}{2} \S_n^{\m\n}\o_{\m\n}$ where $\a_R=\a+\a_5$ and $\a_L=\a-\a_5$ with $\a_5=-\b\mu_5$ characterizes the chirality imbalance (which, for the particle branch of the spectrum, equals to the helicity imbalance). Counting $\a_5$ as $O(\hbar)$, we find the leading contribution of $\a_5$ to be~\cite{Liu:2020flb}
\begin{equation}
\label{spin5}
\begin{split}
\bar{S}_5^\m (\bp) =-\frac{ p^\m\int d\Xi\cdot p \; \a_5\;n_F ( 1 - n_F )}{2E_p\int d\Xi\cdot p\; n_F} ,
\end{split}
\end{equation}
where $p^\m$ is on-shell. A similar term can exist for massive fermions as well, which can be derived by adding a term $-\int d\Xi_\m(y) J_5^\m(y) \a_5(y)$ to the exponent in \eq{rhoLE} with $\a_5(y)$ counted as $O(\pt)$ and following the same procedure in Sec.~\ref{sec:NEDO}~\cite{Becattini:2020xbh}. The observable consequences of $\bar{S}^\m_5$ in heavy-ion collisions have been discussed in Refs.~\cite{Becattini:2020xbh,Gao:2021rom}.

(4) The timelike unit vector $n^\m$ does not need to coincide with $\hat{t}^\m$ (the normal direction of hypersurface $\Xi$) or $u^\m= T\b^\m$ (the flow velocity). We can call $n^\m$ the helicity-frame fixing vector because in its rest frame $\mc{M}^{\m\n}\S^n_{\m\n}$ represents the projection of spin onto the momentum direction. It is crucial to notice that, because the Wigner function is Lorentz covariant, the axial current $\mc{A}^\m(x,p)$ is also Lorentz covariant: $\mc{A}^\m(\L x,\L p)=\L^\m_{~\n}\mc{A}^\n(x,p)$ for an arbitrary Lorentz transformation $\L^\m_{~\n}$. This requires $\mc{A}^\m$ to be actually $n^\m$ independent~\cite{Hidaka:2016yjf,Liu:2020flb,Sheng:2020oqs} and therefore $\o_{\m\n}$ must be $n^\m$ dependent. Under an infinitesimal change $n^\m\ra n'^\m=n^\m+\d n^\n$, the change in $\o^{\m\n}$, $\o_{(n)}^{\m\n}\ra \o_{(n')}^{\m\n}=\o_{(n)}^{\m\n}+\d \o_{(n)}^{\m\n}$ is constrained by the following relation~\footnote{Note that at global equilibrium $\o^{(n)}_{\m\n}$ is determined by thermal vorticity and is thus $n^\m$-independent and \eq{Lorentztransf} identically holds.}:
\begin{eqnarray}
\label{Lorentztransf}
\d\o^{(n)}_{\m\n}\S^{\m\n}_n=\frac{2}{p\cdot n}\S_n^{\m\n}\d n_\n \ls\o^{(n)}_{\m\l}p^\l+\pt_\m(\a+\b\cdot p)\rs.
\end{eqnarray}
In particular, for the massive case, we are allowed to choose $n^\m=p^\m/m$ and the mean spin vector is thus expressed by spin potential only,
\begin{eqnarray}
\label{newmeanspinspecial}
\bar{S}^{\m}(\bp)&=& - \frac{\e_{\m\n\a\b} p^\n\int d\Xi\cdot p \;\o_{(p)}^{\a\b}\;
 n_F ( 1 - n_F )}{8E_p\;\int d\Xi\cdot p\; n_F}.
\end{eqnarray}

(5) In heavy-ion collisions, one may be interested in the global spin polarization $\lan P^\m\ran=\lan\bar{S}^\m\ran/s$ ($s=1/2$ for Dirac fermions) where $\lan\bar{S}^\m\ran$ is the spin vector averaged over whole phase space,
\begin{eqnarray}
\label{globalspin}
\lan\bar{S}^{\m}\ran&=&\frac{1}{N}\int \frac{d^3\bp}{(2\p)^3E_p}\int d\Xi\cdot p \; S^{\m}(x,\bp),
\end{eqnarray}
where $N=2\int\frac{d^3\bp}{(2\p)^3E_p}\int d\Xi\cdot p\; n_F$ is the total particle number. Note that $E_p$ in the denominator of momentum integral is necessary to make the dimension of phase space correct. As a case study, we consider an idealized situation where the normal vector $\hat{t}^\m$ coincides with fluid velocity $u^\m$ and we obtain
\begin{eqnarray}
\label{globalspinspecial}
\lan\bar{S}_{\m}\ran&=&-\frac{1}{4}\e_{\m\n\a\b}\int d\Xi\cdot u \,u^\n \ls\o_{(u)}^{\a\b} F(x)-\frac{2}{3}\D\o_{(u)}^{\a\b} G(x)\rs,
\end{eqnarray}
where $F(x)=N^{-1}\int\frac{d^3\bp}{(2\p)^3}\,n_F (1-n_F)$, $G(x)=N^{-1}\int\frac{d^3\bp}{(2\p)^3}\frac{\bp^2}{E^2_p}\,n_F (1-n_F)$, and we have taken $n^\m=u^\m$ to perform the calculation and used the subscript $(u)$ to indicate that $\o^{\a\b}_{(u)}$ is the spin potential when $n^\m=u^\m$. Note that, in this case, the gradient of $\a$ and the thermal shear do not contribute to the global spin polarization.

\section{Conclusion}
We have re-visited the spin polarization formula for Dirac fermions at local thermal equilibrium. Using two different approaches, the chiral/spin kinetic theory and the method of non-equilibrium density operator, we have derived a Cooper-Frye type formula that connects the mean spin vector (or equivalently, spin polarization vector) of Dirac fermions on a hypersurface with the temperature, chemical potential, flow velocity, and spin potential that characterize the Gibbs state for local equilibrium. The formula is an extension of the previously known results valid at global equilibrium, or under the finite shear tensor, to the situation where a finite independent spin potential is presented. In such a way, the present formula can apply to spin hydrodynamics as machinery to convert the spin potential (and temperature, flow velocity, etc) calculated in spin hydrodynamics to the spin polarization observable on some freeze-out hypersurface. Application to spin hydrodynamics will be our future task. Another future task is to extend the current calculations, which is at $O(\hbar)$ or equivalently at $O(\pt)$ order, to higher orders. Such an extension is straightforward with the chiral/spin kinetic theory and the method of non-equilibrium density operator though the calculations would be more involved.

%

\acknowledgments
We thank Francesco Becattini, Matteo Buzzegoli, Shuai Liu, Shi Pu, Shuzhe Shi, Qun Wang, Yi Yin for useful discussions.
This work is supported by NSFC under Grant No.~12075061 and Shanghai NSF under Grant No.~20ZR1404100. Y.-C.L is also supported by China Postdoctoral Science Foundation under Grant No.~2020M681139 and NSFC under Grant No.~12047516.

{\it Note added.---} During the final preparation of the present paper, we became aware of a simultaneous work~\cite{Buzzegoli:2021wlg} which bears some overlap with our results.

\bibliographystyle{apsrev4-1}
\bibliography{Spin_Polarization_Formula}

\end{document}